# Towards an efficient smart space architecture

Somia Belaidouni
MMS Laboratory, Quebec University,
École de technologie supérieure
Montréal, Canada

Moeiz Miraoui
High institute of applied sciences
and technology
University of Gafsa, Tunisia

Chakib Tadj
MMS Laboratory, Quebec University,
École de technologie supérieure
Montréal, Canada

*Abstract*—A smart space offers entirely new opportunities for end users by adapting services accordingly to make life easy. A number of architectural designs have been proposed to design context awareness systems and adaptation behavior. However, the quality of the system depends on the degree of satisfaction of the initials needs. In this paper, we discuss three main indicators of quality design for smart spaces that are strongly related to the context modules and reasoning process: functionality, reusability and changeability. A general layered architecture system is presented to define the principal components that should constitute any context aware adaptive system.

## I. INTRODUCTION

A smart environment can be defined as a space that is able to acquire and apply knowledge about its environment and inhabitants to provide appropriate services [1]. Some important features of smart environments are that they possess a degree of autonomy, adapt themselves to changing environments and communicate with humans in a natural way [23]. This ubiquitous environment is composed of different devices such as embedded computers, multimodal sensors and various software programs that aim to facilitate human life by offering abundant information from different devices to accomplish the required expectations. The capacity to build a system that is able to process the information delivered to meet the user's requirements is dependent on an architecture that can support the dynamism and heterogeneous devices and handle the interactions between the user and system. Therefore, the main objective of any architecture developed for any smart space is the ability to sensor and gather information from the area where it is deployed and to process it by making adequate decisions and executing actions on the physical environment.

The majority of the existing architectures for smart space focus on acquiring data from sensors, interpreting the data and adapting services to adequately assist users to concentrate on their specific tasks. Context awareness and adaptation are tightly related, and the two terms are often used as synonyms [26]. However, adaptation means the ability to change a service and produce another corresponding environment; context awareness is the ability to perceive the different situations of users to adapt actions before execution. The proposed architectures are designed at a limited concentration at the modularity of components. Moreover, they do not take into account the abstraction of context received from the environment. In addition, the awareness of context and the ability of adaptation by using reliable algorithms still remain to be research questions. These different problems involve development complexity in making the system more difficult to maintain and to reuse components.

The main goal of this paper is to define the most important quality characteristic related to a smart system. A set of indicators is introduced to reveal the highest requirements in a height quality system. Subsequently, we present a general architecture with essential components that can be designed to build an adaptable system with high performance. We then present a comparative study between our proposal design and other attractive structures based mainly on the evaluation criteria.

The rest of this paper is structured as follows. In the next section, we review related work. Section 3 identifies three essentials metrics for creating a better architecture for a smart space. Section 4 provides an overview of the proposed architecture and all of the components necessary for building an adaptable system. In section 5, we analyze and compare our proposed design with others systems, and section 6 concludes the paper.

## II. Related Work

Researchers studying smart spaces have been actively doing research in recent years, as many related smart systems have been developed and tested in the real world. The main goal for any smart system is to adapt compartments according to the user's preferences and satisfy all demands.

In this section, we showcase some smart environment projects as case studies and discuss them. Our studies are based on projects and not necessarily on awareness context or self-adaptive systems. However, we focus on reliable projects that have proven feasibility and validity. We also aim to study distinct structures with different modules, where some modules can be combined or removed altogether according to their functionalities. We can thus expand our performance analysis and deduct the primary criteria that affect the working of the design. The MavHome project [7] aims to create a home that acts like a rational agent. The basic idea is to build a smart home with sensors and actors to produce services adapted to the inhabitants' activities. In the proposed solution, the main idea is to use agents that work together for selecting and initializing the adequate action and hence maximizing the goals for the home. Due to this reason, the design home must be able to receive and process the information and perform reasoning to adapt actions to user preferences. The MavHome architecture is structured in a hierarchical form composed of agents. Each agent is separated into four layers. The physical layer contains all of the hardware that may exist in the home, including sensors, actioners and hardware for communication.

The communication layer has the responsibility of transmitting information, queries or requests between agents. The information layer gathers and stores any data in the system and generates knowledge useful for making the appropriate decisions. The decision layer is based on the previous layer information for making decisions and selecting related actions. The SOCAM project [4] is an architecture developed for a service-oriented context-aware middleware to provide context-aware mobile services within a smart space. The SOCAM architecture is composed mainly of the following components: context provider, context interpreter, context knowledge and context reasoner, service locating service, context- aware mobile service and context database. Context providers provide internal and external context at higher orders, which can be used directly by services. The context interpreter is responsible for context processing and defining the reasoning component based on ontologies that allow for robust approach reasoning. The principal task of ontologies may include deriving high-level contexts from low-level contexts, maintaining the consistency of context knowledge and resolving context conflicts [24]. The context-aware services refer to all elements that use context at different levels, such as agents, applications and services. This component is able to obtain the context of interest by querying the service-locating service to locate all of the context providers, which is the source context. The service locating service deploys multiple services to allow users and applications to discover contexts, locate context providers and adapt to their dynamic changes. The SmartLab project [12] is an OSGI-based middleware platform that provides a cooperative and programmable intelligent environment. The interest in the project stems from the fact that it is an evolutionary and self-configuration structure that adapts services according to rapid changes in such environments. The SmartLab paradigm has already been used as a base for diverse projects. The sensing and actuation layer is composed of a set of devices that constitute the environment. They include the EIB/KNX automation bus, VoIP and VideoIP, indoor location systems, Smart Display, and Smart container. The next layer is the service abstraction layer that interprets and transforms the functionality of the devices received from the previous layers into software services.

The Semantic Context Modeling and Service management, also performed by the server SmartLab, acts like a gateway between the advanced OSGI services and the functionalities of the equipment deployed in this environment. It comprises the Service Manager, which handles the activation and deactivation of devices and, thus, service availability. The Semantic Context Manager stocks knowledge on devices using a common ontology. The final layer is the Programming, Management and Interaction layer, which enables applications to make use of the SmartLab infrastructure. It also generates two generic applications, Environment controller and Context-Manager Front End, which ensure the control and management of an intelligent environment. CoBra [18] is an architecture and specification based on agents for creating, distributing and managing context-aware applications in an intelligent space. The core of any CoBra-distributed system is the Object Request Broker (ORB). The role of the ORB is to maintain a shared model of context between whole entities in the smart space. In addition, the ORB is able to ensure communication between components and heed to issues related to distribution and heterogeneity of the environment.

The concept of the CoBra architecture is based on four main functionalities: context knowledge, context reasoner engine, context acquisition module and privacy management module. Context knowledge uses ontologies for the management and storage of context data. It also provides a set of APIs that allow access to the data context. The context reasoner engine uses a set of ontologies to reason and infer new rules that can be used by the system. It allows also for a more efficient reasoning and sharing of contextual information. The Gator Tech Smart House (GTSH) [9] is a project developed to construct an experimental laboratory and an actual live-in trial environment to test and prove the performance of technologies and systems developed for smart spaces. The smart house is designed to assist older persons or individuals with special needs to maximize their independence and provide a better life for them. The project architecture is developed in six main layers, and the middleware is the main unit connecting the other entities. The physical layer gathers various devices and appliances in the house. It includes sensors and actuators and can also include equipment such as the TV, a lamp, a doorbell or sensors and actioners. The sensor layer communicates with different devices in the physical layer, receives their information and transforms it into useful information to relay from the physical layer into software services that can be used in other services invested into by the next layer. The service layer comprises the central middleware OSGI, which maintains and controls the service bundle.

The middleware store service bundle definitions for each sensor or captor represented in the physical layer. The services are composed of other services that represent the many applications available in the smart space. The knowledge layer represents the ontology mechanism offered for all services and appliances in the house. This allows for reasoning related to the services used. The context management layer is responsible for defining and eliminating the context of interest to be used by applications. The context is designed as a graphical OSGI bundle linking various sensors and captors. The application layer is the last layer equipped with various tools for managing applications and developing the smart environment, and it is therefore responsible for activating or disabling service execution according to the inhabitant's need. The Smart-M3 NOKIA platform [11] is designed for creating a simple, extensible and interoperable model to be deployed on top of any existing infrastructure in a smart environment. Furthermore, it allows devices to easily share and access local and global semantic information. It mainly consists of two components: Semantic Information Broker (SIB) and knowledge processor (KP). The SIB component maintains and stocks all information by exploiting the Resource Description Framework (RDF).

There may be one or many SIBs in the smart space, and they are connected by the protocol SSAP. They work together to manage and provide the same information for any device. KP entities are deployed by developers on different devices participating in the smart space and interact with each other to access SIB for reading or publishing data via the SSAP protocol. Each KP has a specific task, and devices may contain diverse KPs with different functions. However, KPs do not have the ability to communicate with others; instead, they must pass through the SIB unit. The Smart Space Access Protocol (SSAP) is the protocol used by KBs to access a SIB. It performs seven major operations that specify actions the KB or SIB should take. SSAP is session-based, that is, when a KP wants to communicate with the SIB, it must open and active a SSAP session. For this purpose, development of KPs and SIB should support all SSAP operations. This will guarantee interoperability in a heterogeneous environment according to the requirement for smart space. The framework S2CAS (Smart Space Context Aware System) [19] is developed to manage context and to provide services that take into account environmental data. The overall project is designed in three major layers.

The first layer, which is situated at the bottom of the system, can receive information from different sensors and devices and prepare them for the core layer. The layer is equipped with a mechanism for pre-treatment of the captured data. The context information should be physical or virtual data. The second layer is the core of the system. It includes three essentials modules that realize the interpretation, treatment and management of context: the context information processing, the context explanation and the context manager. Context information processing involves receiving information from the previous layer and transforming it to more formal data. The context interpreter, which is composed of a context knowledge base and the context inference engine, records the historical context environment and transforms the low-layer context information into high-layer information. The top layer is defined as the interface between the user and system and presents applications and services according to needs. This layer uses the context height level from the interpreter module of the core layer to provide specific context-aware service.

The Greenerbuilding project [8] is a service-oriented approach (SOA) for designing and developing a building management system. It is a structure that is dedicated to creating an intelligent office in a green and energy-efficient way while preserving and maintaining comfort for habitation. At the same time, it allows its users to modify and adjust the rules of the system for adapting services according to their needs. A significant component of this project involves using renewable energy, which is an excellent alternative for economic and environmental reasons. In addition, with the introduction of the SOA principles, the building can take into account the heterogeneity of the available devices and capabilities. The physical layer consists of various devices in the building, which can include sensors, actuators, gateways and also low-level protocols. The gateway sensors and actuators collect all information from the devices and transfer it to the height level of the system. The Greenerbuilding is interconnected to smart grid services that make it aware of external and internal energy pricing. This allows Greenerbuilding to adjust demands and reduce the energy costs according to the operations. A ubiquitous layer ensures the functioning of the whole system. It is composed of a repository that contains a database of the devices and their configurations and states. The context component collects information from the sensors and transfers it to the height level. The composition layer is the last layer that supplies the control services and represents the system's interface to its users. It mainly performs reasoning and makes decisions using components based on the Rule Maintenance Engine. It gathers information about user preferences makes decision and decides on actions to be performed according to user requirements. Self-adaptation systems [6] are a new project designed to develop an innovative approach for building, running and managing a self-adaptation system for smart space. The main idea is to make smart and reliable devices handle uncertainty and uncontrolled conditions in smart spaces. It implements agents in such a way that each agent defines a device, and all agents cooperate to provide seamlessly adequate services for users. The system consists of adaptive software and an adaptation agent. The adaptive software consists of a SAManager thread, which is the central component of the system. It contains the predefined normal status constraints (NSCs) and function list to be executed. SAManager performs multiple duties. Initially, it is invoked by the adaptable system to create the first thread. It then monitors and manages the NSC list. Moreover, SAManager is responsible for selecting functions related to adaptation commands from the adaptation agent and notifying that an anomaly occurred.

The adaptation agent includes the adaptation process and logics in contact with the SAManager, and it states its demands and selects the appropriate adaptation strategy using the adaptation strategy table (AST). The adaptation agent receives data from different resources and may collaborate with neighboring adaptation agents to provide services. Thus, the implementation of the adaptation agent can be customized and configured according to the goals of the system.

### III. SMART SPACE REQUIREMENTS

It is widely acknowledged that a smart space is a typical open, distributed and heterogeneous pervasive computing system, which aims to create a ubiquitous, human-centric environment with embedded computers, information appliances and multimodal sensors that facilitate humans to perform tasks efficiently by offering abundant information and assistance from computers [32]. A prominent and important requirement for a smart space to be able to respond adequately to user's needs is to assure interoperability between components. Interoperability reflects the existence of a co-operative structure of services [31], which means the capability to mask the heterogeneity of the equipment to facilitate communication within the environment. A smart space is a grouping of heterogeneous entities (sensors, objects, devices) that interact and service one another to complete different actions. Scalability is another important requirement for designing a smart space. A reliable system designed for a smart space should enable its structures and functions [33]. A smart system should also adapt the behavior and the interface of an application to the user situation and equipment to provide information about a physical environment, which is shared with inherently dynamic applications [34]. In addition, the concept of context is the principal entry on which all of the system processes are based. It denotes various aspects such as where one is, who one is with, what one is doing and what resources are nearby (in other words, location, identity of neighbor entities, activity and the state of the physical environment) [35]. The ability to adapt services in such an environment requires some technologies for sensing, acquiring, interpreting and inferring context, which is the principal entry to the smart system. This mechanism is known as awareness of context and is an essential step towards improving the quality of the system.

### IV. QUALITY ARCHITECTURE

The quality of an architecture designed for a smart system is measured by the ability to achieve the planned targets [1]. Thus, height quality must respond to the growing demands of users in a reliable and error-free way. To evaluate the smart space architecture quality, different quality characteristics might be of interest. In the context of this paper, we define three basic characteristics: functionality, reusability and changeability

- Architecture functionality (Q1) - the architecture of the system satisfies the stated objective of the system.
- Architecture reusability (Q2) - can the architecture (or components) of the system can be reused in other systems?
- Architecture maintainability (Q3) - can the architecture of the system be maintained to extend and adapt to a changed environment?

We choose the architecture functionality characteristic because it is strongly related to the goals of any smart space to provide services for the tasks. Furthermore, we choose reusability and maintainability, the ISO/IEC9126 sub-characteristics of maintainability because we consider them to be two important concepts that should considered by the architecture.

We further refine architecture functionality into two concrete quality characteristics:
- Awareness context: this term signifies that the system adapts its services to a user's preferences and activities in the environment. The awareness context is becoming increasingly more prevalent and can found in many areas of research [28] due to its importance. In a smart environment, a context awareness system must have the ability to detect, sense, interpret and respond to user requests [20]. Thus, this property leads to detection of contextual information from different sensors or device sources but also to understanding what type of contextual information should be provided for the appropriate service.

- Self-adaptation: the capability of the system to adjust its behavior in response to the environment in the form of self-adaptation has become one of the most promising research directions [2]. We consider that a self-adaptation architecture is capable of readjusting its behavior to external conditions. It is widely known that a smart space is richly equipped by different sensors/devices and connected to diverse technologies. Within this heterogonous environment, applications must modify and adapt their services to the user's needs according to the current context [14]. Therefore, we can say that an adaptive system is a context-awareness system that applies specific adaptation approaches. Moreover, a smart system is characterized by self-adaptation that does not require any tiers. However, the system triggers its attitude automatically according to the actual context.

We refine architecture reusability and maintainability in the modularity that provides the mechanism for separating the complex system into functional modules. The modularity must be effective: each module should focus exclusively on one aspect of the system, or the modules should be cohesive. Additionally, modules should be independent of each other. A more efficient modularity leads to a more understandable, reusable and modifiable architecture. The maintainability of a software system is the ease with which it can be modified to changes in the environment, requirements or functional specifications [13]. The maintainability process involves the capability of making modifications at different levels of conception. Adaptive systems designed for smart spaces are to be modified several times after their initial development, taking into account the dynamism of the environment, to facilitate system performance. A number of studies [27] have shown that 50% to 70% of the total lifecycle cost of a software system is spent after initial development due to the high cost of software maintenance. This propriety defines an important quality that we should consider for fulfilling system purposes. We now study some quality indicators that can be used to check the achievement of these characteristics.

We define two main indicators that can be measured for the awareness of context:
- Abstraction of context
- Modification of context

The abstraction of context aims to model and present real-world situations. Information from physical sensors is called low-level context and is acquired without any further interpretation [29]. Moreover, human interactions and behavior are received in unreadable format. This limitation of low-level contextual cues risks reducing the usefulness of context-aware applications [30]. One way of overcoming this problem is to derive higher-level context information from raw sensor values, based on context reasoning and interpretation. The main idea of the abstraction process is to create a model that takes raw sensor data as input and generates more significant data relevant to the current situation.

The identification of components and their relationships is the essence of high-level software design [3]. The modularity of the architecture can be measured by the degree of coherence and coupling [17]. These measures capture the degree of interaction and relationships between the elements [22]. Software research shows that designs with low coupling and high cohesion enable the production of a more reliable and maintainable system [10]. Principally, cohesion refers to the degree of interconnection within the same module. High cohesion of a module indicates that the elements are interconnected well for implementing a specific function together, without outside assistance. The coupling defines the degree of interdependence between the modules. High coupling of a module creates a strong dependency with other modules, which means that its functionality is based heavily on the presence of other modules. Figure 1 shows the principal characteristics and the relationship between indicators to produce a high-quality architecture.

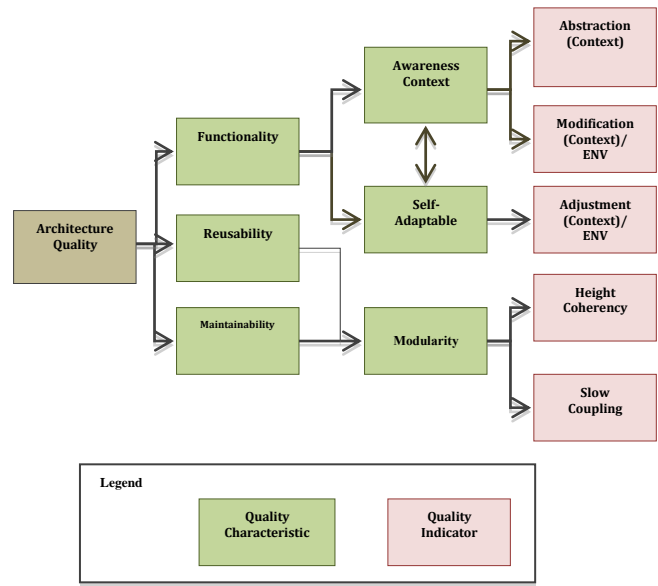

Figure 1. Quality architecture for smart space

## V. Architecture Overview

In this section, we will present the design pattern for adapting services in a smart environment. Using the most successful projects as case studies, we searched for a common pattern and advantageous structure for building a smart system. We wanted to provide a general design for adapting actions according to the different versions of context in a smart space. This pattern may be used as an exemplar for other projects and is intended to reduce efforts and allow developers to concentrate just on specifying requirements tailored to every project. Figure 2 presents the pattern process in an entry system context (such as battery level and network resources) and for users (such as localization, noise, and user needs). The whole pattern may be split into four layers (physical, context, adaptation, decision) in such a way that every layer is a set of different components that interact to achieve specific services that will be presented to the user at the end. It is important to note that all components of a layer must collaborate to generate services to the following layer.

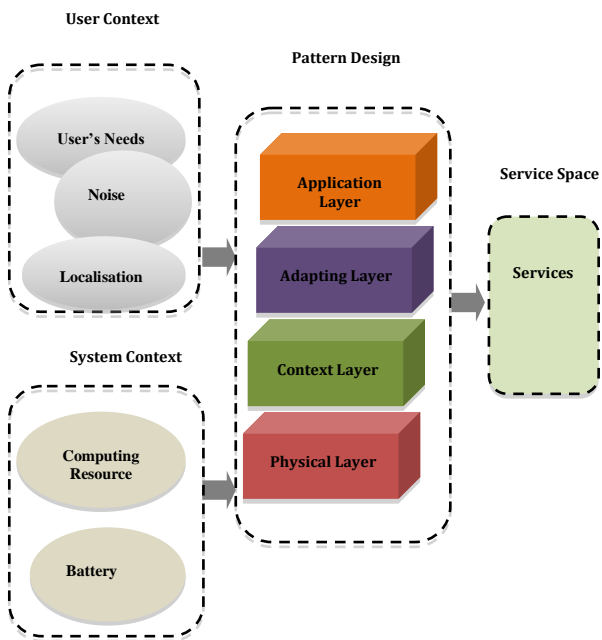

Figure 2. Design pattern for a smart system

## VI. General Architecture for Smart Spaces

In this section, we will present a generic design of the smart environment architecture. We intend to provide a quality architecture that satisfies the requirements mentioned earlier. Therefore, our architecture includes four layers with several distinct components in every layer. It is important to note that these components are not exhaustive in terms of availability in the system. However, projects may have a greater or smaller number of components, depending on the goals and the environment. The general architecture is presented in Figure 3.

### A. Physical Layer

Is a wireless network of smart objects. It contains all of the hardware parts of the system, which includes wired and wireless sensors and actuators associated with their infrastructure. The main goal of this layer is to capture and collect external information about the environment and transfer it to the next layer.

### B. Context Layer

Is the backbone of the whole architecture. It is the main support of the system that contains several components with different responsibilities related to the context. We can split this layer into four essential modules that reside in every smart environment in one form or another.

#### a) Knowledge Base

The knowledge base is the base of any information related to the system and its components. It can include static information about devices, appliances or users such as configuration information or dynamic information exposed to frequent changes, such as the location of the users or devices in the environment. Moreover, this module is a base of all information required for high-level reasoning. The nature of the data depends on the type of reasoning module. For an ontology-based system, this information would represent ontologies, and for a rule-based system, the reasoning rules would be stored.

#### b) Context Module

The main responsibility of this module is to transform low-level data gathered from the previous layer into higher-level information. This module collects sensory data and interprets them to identify the meaning and yield understandable data that are usable by the reasoning module. The reasoning module includes other components that work together to manage the data.

#### c) Aggregators

Aggregators collect different pieces of context information that are logically related in a logical depot. A context is often received from different and distributed sensors. Therefore, aggregators decompose and gather context information in coherent groups for context inference, consistency checking and knowledge sharing.

#### d) Interpreter

After combining context information in logical groups, the interpreter will identify the meaning of the context before it can be used. An application may not be interested in the low-level data. However, it requires comprehensible data that are expressed at higher levels of abstraction.

e) *Reasoning Procedure*

Context reasoning, also known as context inference, is one of the most important elements in the context aware system and involves different process: checking the consistency of the context and deducing high-level, implicit context from low-level, explicit context [5]. For achieving these goals, the reasoning process is based on specific reasoning methods such as rules, fuzzy logic, decision trees and Bayes networks. For our work, we choose the ontology mechanism because it can exploit various existing logic mechanisms to deduce high-level conceptual context from low-level, raw context and check for and solve inconsistent context knowledge due to imperfect sensing [2].

f) *Modeling*

Context modeling is an important instrument for handling contexts and managing how they are collected, organized and represented, whereas context awareness signifies reasoning about the context [36]. In a smart space, context modeling defines the manner for representing context information for reasoning and adaptation processes. Basically, it aims to provide more simplified descriptions using different mechanisms. There are different models for present context such as the following: graphical models, logical models and ontology-based models. It is important to note that researchers show that an ontology is the most advantageous and beneficial model because it simultaneously allows for recognizing the context being operated in and reasoning about other different contexts. Furthermore, using non-ontology-based models requires much programming effort and tightly couples the context model to the rest of the system [16].

C. *Adaptation Layer*

It comprises the domain-level logic of the system. It aims to adapt services according to the context detected by the preceding layer. It is responsible for analyzing context and deciding upon actions to execute. Such an adaptation is required to serve several goals determined by the nature of the application, the nature of the context and the preferences of the user [15]. According to its responsibilities and definitions, we summarized this layer into six essential processes.

a) *Machine Learning*

Machine learning is a type of automatic learning that aims to provide the best possible decision and action to execute. In smart systems, machine learning is applied to support reasoning and inference related to complex information. As defined in [21], machine learning in a pervasive system is an innovative approach that integrates wireless, mobile and context-awareness technologies to detect the context of learners in the real world and thus accordingly provide adaptive support, personalized services or guidance.

It integrates wireless, mobile and context-awareness technologies to detect the situation. This process is also linked to database user preferences to consider user preferences during the learning process. Several algorithms are employed depending on the project goals and its requirements, such as genetic algorithms, artificial neural networks, decision trees and reinforced learning.

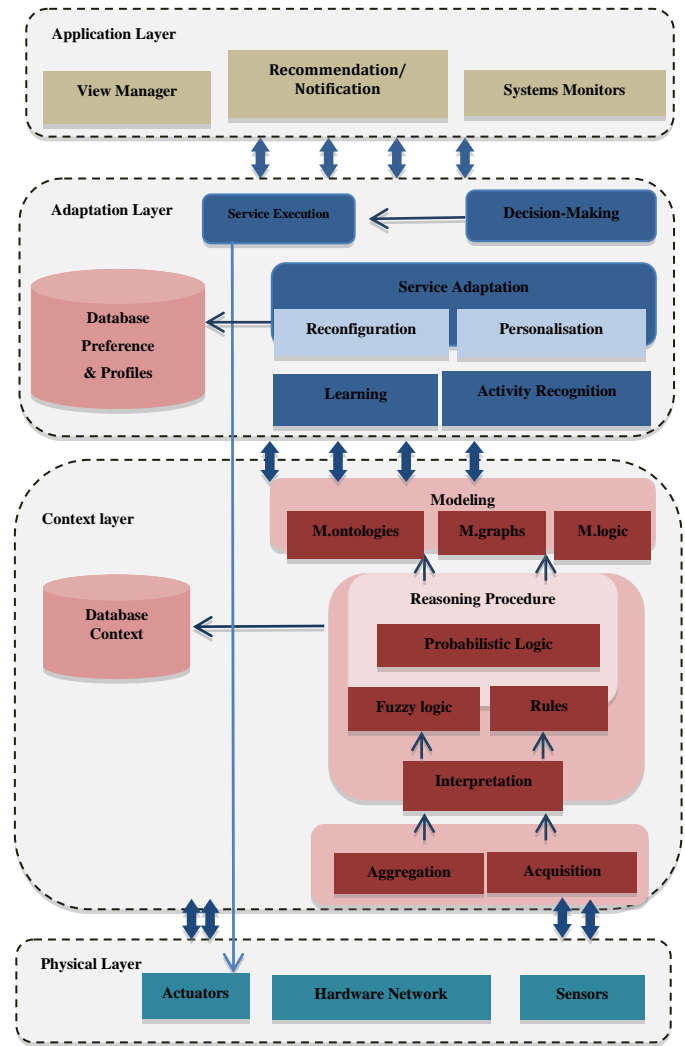

Figure 3. General architecture for a smart space

b) Activity Recognition

Activity recognition is a process that uses context information about the current state and applies internal knowledge to clarify the data. This activity should be performed at discrete time points in real time in a progressive manner [25]. This activity is able to detect a current service or movement based on the information received by sensors to recognize the current state of the ongoing activity and further identify the user's needs.

### c) Adaptation Services

Adaptation tools form the core of a context aware adaptive system. The input of this engine is the learning context, and the output of this component is the adaptation action. The adaptation process is related to learning activities that add adjustments to meet user needs. The process adaptation can be used similar to configuration, personalization or recommendation.

### d) Decision-Making

Decision-making is responsible for deciding adequate actions that should be performed for a given context information. This process receives queries from other modules to decide which action will be performed. In effect, it is quite a delicate module because it is based mainly on the received information and it must understand, model and learn information well to provide a good decision.

### D. Application Layer

It is the upper layer of the system that supports the implementation of services. Principally, this layer is composed of three parts: (1) view manager that specifies mechanisms that manage the diverse views and is responsible for handling the different interfaces for receiving external context, displaying the user interface and managing the user interaction. (2) Recommendation/notification is an adaptive application behavior that describes the process of reasoning and adaptation for every executed action to understand the situation and the respective services. It is important to note that this component receives all of the information from the adaptable layer. (3) The system monitors that include all information related to the systems and their performance indicators, status of components and any detected errors from inside the system or environmental devices.

## VII. Performance Analysis

The goal of the performance analysis of any architecture is to verify that the implemented system meets all of the requirements for the current project. In this section, we aim to compare our proposal architecture with seven systems developed for adapting services in a smart space. To conduct our analysis, we focus on three criteria mentioned subsequently: design of the system and the contextual and reasoning modules. These three characteristics are unique for every design and differ according to the environment and the project goals. Table 7 shows a comparison of the relevant smart systems.

The design of the system reflects the nature of interconnection and relationship between the modules and their elements. The ideal architecture intends to minimize coupling (coup) and maximize cohesion (coh) to reduce the efforts related to understanding and maintaining the systems. By analyzing related work, we note that several systems are highly cohesive and coupled. Such drawbacks make the system complex, inflexible and ineffective. SOCAM and Greenbuilding are two modular architectures that respect a high cohesion and a low coupling. This quality facilitates the changeability and maintainability of systems. The abstraction of the context and context awareness are major criteria for processing and identifying more comprehensible data to provide intelligent and adaptable context-aware services. MavHome, CoBra and SOCAM present the advantages of being aware of the context and providing a level of abstraction during context processing. Reasoning module is very important, especially in a smart space, where the context is received from sensors at low levels. The reasoning engines have the capability of deriving new concepts to adapt service behavior accordingly. We remark that the CoBra, SOCAM, Smart-M3, Smartlab systems use ontologies to reason and provide a rich formalism for specifying contextual information.

By comparing the other systems with our proposed system, we note that our solution responds effectively to all of the requirements (components modularity, contextual/reasoning module), unlike the other solutions we surveyed. A further advantage is that our solution involves a full architecture that contains all of the layers and components for managing the context and responding to user requests. In addition, our solution embeds reasoning and adaptation processes to regulate actions dynamically according to the environment and its users.

Table 7. Comparison of smart systems

| Systems | Component modularity | | Contextuel module | | Reasoning module | |
|---|---|---|---|---|---|---|
| | Coh | Coup | Context abstraction | Context awareness | Reasoning module | Type of reasoning |
| MavHome | High | High | + | + | + | Machine learning |
| CoBra | Low | High | + | + | + | Ontologies |
| SOCAM | High | Low | + | + | + | Ontologies |
| Self-adaptive | High | High | + | - | - | - |
| Smart-M3 | High | High | - | + | + | Ontologies |
| S2CAS | High | High | + | - | + | Rules |
| Green-building | High | Low | + | - | + | Rules |
| SmartLab | High | High | + | - | + | Ontologies |

VIII. CONCLUSION

The development of adaptable systems has undergone an important evolution marked by an incremental improvement in techniques for managing and reasoning related to the context.
In this paper, we present the most important characteristics for quality architecture and the requirements necessary for developing a reliable adaptable system: functionality, maintainability and reliability. We subsequently present general-layer architecture with the most necessary components to develop a smart space. From the analysis of some relevant architecture, we show that the modularity of components, awareness of context and process reasoning are three important characteristics that reflect the reliability of systems.

Future development efforts should focus more on the development of modular architectures with high cohesion and low coupling components that are important for facilitating the reusability and extensibility of systems. Also, an efficient model for handling context data is essential for the development of such systems. To achieve this, we should concentrate on how to model context and establish its elements in a clear and precise manner. On the other hand, context inference is a hard task and requires methods to reason correctly and adapt systems automatically to new situations. We must therefore reinforce the development of more efficient reasoning methods that take into account the dynamism of smart spaces and their entities. Security and privacy are important parameters, which we must consider in the adaptive system to protect sensitive information.

REFERENCES

[1] Handte, Marcus, et al. "3PC: System Support for Adaptive Peer-to-Peer Pervasive Computing." ACM Transactions on Autonomous and Adaptive Systems (TAAS) 7.1 (2012): 10.
[2] Brun, Yuriy, et al. "Engineering Self-Adaptive Systems Through Feedback Loops." Software Engineering for Self-Adaptive Systems. Springer Berlin Heidelberg, 2009. 48-70.
[3] Allen, Edward B., Taghi M. Khoshgoftaar, and Ye Chen. "Measuring Coupling and Cohesion of Software Modules: an Information-Theory Approach." Software Metrics Symposium, 20ETRICS 2001. Proceedings. Seventh International. IEEE, 2001.
[4] Gu, Tao, Hung Keng Pung, and Da Qing Zhang. "A Middleware for Building Context-Aware Mobile Services." Vehicular Technology Conference, 2004. VTC 2004-Spring. 2004 IEEE 59th. Vol. 5. IEEE, 2004.
[5] Wang, Xiao Hang, et al. "Ontology Based Context Modeling and Reasoning Using OWL." Pervasive Computing and Communications Workshops, 2004. Proceedings of the Second IEEE Annual Conference on. IEEE, 2004.
[6] Chun, Ingeol, et al. "An Agent-Based Self-Adaptation Architecture for Implementing Smart Devices in Smart Space." Telecommunication Systems 52.4 (2013): 2335-2346.
[7] Cook, D.J., Youngblood, M., Heierman, E., Gopalratnam, K., Rao, S., Litvin, A., Khawaja, F. "MavHome: An agent-based smart home", in Proceedings of PerCom 2003, 521-524, 2003.
[8] Degeler, Viktoriya, et al. "Service-oriented architecture for smart environments (short paper)." Service-Oriented Computing and Applications (SOCA), 2013 IEEE 6th International Conference on. IEEE, 2013.
[9] Helal, Sumi, et al. "The Gator Tech Smart House: A programmable Pervasive Space." Computer 38.3 (2005): 50-60.
[10] Holvitie, Johannes, and Ville Leppanen. "Illustrating Software Modifiability--Capturing Cohesion and Coupling in a Force-Optimized Graph." Computer and Information Technology (CIT), 2014 IEEE International Conference on. IEEE, 2014.
[11] Honkola, Jukka, et al. "Smart-M3 Information Sharing Platform." The IEEE symposium on Computers and Communications. IEEE, 2010.
[12] Lopez-de-Ipina, Diego, et al. "Dynamic Discovery and Semantic Reasoning for Next Generation Intelligent Environments." (2008): 23-23.
[13] Bengtsson, Per Olof, et al. "Analyzing Software Architectures for Modifiability", Blekinge Institute of Technology Research Report 2000:11, ISSN: 1103-1581.
[14] Miraoui, Moeiz, et al. "Context-Aware Services Adaptation for a Smart Living Room." Computer Applications & Research (WSCAR), 2014 World Symposium on. IEEE, 2014.
[15] Mizouni, Rabeb, et al. "A Framework for Context-Aware Self-Adaptive Mobile Applications SPL." Expert Systems with Applications 41.16 (2014): 7549-7564..
[16] Niemelä, Eila, and Juhani Latvakoski. "Survey of Requirements and Solutions for Ubiquitous Software." Proceedings of the 3rd International Conference on Mobile and Ubiquitous Multimedia. ACM, 2004. Conference on Mobile and Ubiquitous Multimedia, College Park, Maryland, USA, 2004.
[17] Stevens, Wayne P., Glenford J. Myers, and Larry L. Constantine. "Structured design." IBM Systems Journal 13.2 (1974): 115-139.
[18] Sukalikar, Shriya, Sudhakar Kumar, and Niyati Baliyan. "Analysing Cohesion and Coupling for Modular Ontologies." Advances in Computing,

Communications and Informatics (ICACCI, 2014 International Conference on. IEEE, 2014.

[19] Sukalikar, Shriya, Sudhakar Kumar, and Niyati Baliyan. "Analysing Cohesion and Coupling for Modular Ontologies." Advances in Computing, Communications and Informatics (ICACCI, 2014 International Conference on. IEEE, 2014.

[20] Jiang, Zhen Ming, Ahmed E. Hassan, and Richard C. Holt. "Visualizing Clone Cohesion and Coupling." Software Engineering Conference, 2006. APSEC 2006. 13th Asia Pacific. IEEE, 2006.

[21] Wang, Shu-Lin, and Chun-Yi Wu. "Application of Context-Aware and Personalized Recommendation to Implement an Adaptive Ubiquitous Learning System." Expert Systems with Applications 38.9 (2011): 10831-10838.

[22] Újházi, Béla, et al. "New Conceptual Coupling and Cohesion Metrics for Object-Oriented Systems." Source Code Analysis and Manipulation (SCAM), 2010 10th IEEE Working Conference on. IEEE, 2010.

[23] Das, Sajal K., and Diane J. Cook. "Designing Smart Environments: A Paradigm Based on Learning and Prediction." Pattern Recognition and Machine Intelligence. Springer Berlin Heidelberg, 2005. 80-90.

[24] Gu, Tao, Hung Keng Pung, and Da Qing Zhang. "A Middleware for Building Context-Aware Mobile Services." Vehicular Technology Conference, 2004. VTC 2004-Spring. 2004 IEEE 59th. Vol. 5. IEEE, 2004.

[25] Chen, Liming, Chris D. Nugent, and Hui Wang. "A Knowledge-Driven Approach to Activity Recognition in Smart Homes." Knowledge and Data Engineering, IEEE Transactions on 24.6 (2012): 961-974.

[26] Chaari, Tarak, Frédérique Laforest, and A. Celentano. "Design of Context-Aware Applications Based on Web Services." INSA Lyon, France, Tech. Rep. RR-2004- 033 (2004).

[27] Hoefler, Dorothea, et al. "Software Maintenance Management." U.S. Patent No. 8,176,483. 8 May 2012.

[28] Krumm, John,. Ubiquitous Computing Fundamentals. CRC Press, 2009.

[29] Ye, Juan, et al. "Using Situation Lattices to Model and Reason about Context." Proceedings of MRC 2007 (coexist with CONTEXT'07) (2007): 1-12.

[30] Bettini, Claudio, et al. "A Survey of Context Modelling and Reasoning Techniques." Pervasive and Mobile Computing 6.2 (2010): 161-180.

[31] van der Meer, Sven, et al. "Design Principles for Smart Space Management." 1st International Workshop on Managing Ubiquitous Communications and Services (MUCS). 2003.

[32] Qin, Weijun, Yue Suo, and Yuanchun Shi. "Camps: A Middleware for Providing Context-Aware Services for Smart Space." Advances in Grid and Pervasive Computing. Springer Berlin Heidelberg, 2006. 644-653.

[33] Wu, Zhaohui, et al. "ScudWare: A Semantic and Adaptive Middleware Platform for Smart Vehicle Space." Intelligent Transportation Systems, IEEE Transactions on 8.1 (2007): 121-132.

[34] Chaari, Tarak, Frédérique Laforest, and A. Celentano. "Design of Context-Aware Applications Based on Web Services." INSA Lyon, France, Tech. Rep. RR-2004-033 (2004).

[35] Bucur, Doina. "On Context Awareness in Ubiquitous Computing." Diss. Aarhus University, Det Naturvidenskabelige FakultetFaculty of Science, Datalogisk Institut, Department of Computer Science.

[36] Topcu, Ferit. "Context Modeling and Reasoning Techniques." SNET Seminar in the ST. 2011.